\documentclass[11pt,a4paper]{article}
\usepackage[a4paper,total={160mm,220mm}]{geometry}
\usepackage{amsmath,amssymb}
\usepackage{cite}
\usepackage{braket}
\usepackage{hyperref}
\usepackage[capitalise]{cleveref}

\newcommand{\abs}[1]{\left\lvert{#1}\right\rvert}

\newcommand{\md}{\mathrm{d}}
\newcommand{\me}{\mathrm{e}}
\newcommand{\vol}{\mathrm{vol}}

\begin{document}
\numberwithin{equation}{section}
\title{
\vspace*{-0.5cm}{\scriptsize \mbox{}\hfill MITP-23-034}\\
\vspace{3.5cm}
\Large{\textbf{Quantum Imprint of the Anharmonic Oscillator}}
\vspace{0.5cm}}

\author{Prisco Lo Chiatto, Sebastian Schenk, and Felix Yu\\[2ex]
\small{\em PRISMA$^+$ Cluster of Excellence \& Mainz Institute for Theoretical Physics,} \\
\small{\em Johannes Gutenberg-Universit\"at Mainz, 55099 Mainz, Germany}\\[0.8ex]}

\date{}
\maketitle

\begin{abstract}
\noindent
We study the anharmonic double well in quantum mechanics using exact Wentzel-Kramers-Brillouin (WKB) methods in a 't~Hooft-like double scaling limit where classical behavior is expected to dominate.
We compute the tunneling action in this double scaling limit, and compare it to the transition amplitude from the vacuum to a highly excited state.
Our results, exact in the semiclassical limit, show that the two expressions 
coincide, apart from an irreducible and surprising instanton contribution.
Thus, the semiclassical limit of the anharmonic oscillator betrays its quantum origin as a rule, which we dub the ``quantum imprint rule," showing that the quantum theory is intrinsically gapped from classical behavior.
Besides an example of the failure of reductionism and an example of a resurgent connection between perturbative and nonperturbative physics, this work provides a possible classification of theories according to their quantum imprints.
\end{abstract}

\newpage

\section{Introduction}
\label{sec:introduction}

A central puzzle in the study of quantum phenomena is the possible emergence of a classical limit.
Empirically, of course, classical physics dominates everyday interactions while the fundamental quantumness of Nature has been established to extraordinary precision~\cite{FeynmanLectures3}.
The simplest explanation for this disparate behavior is that the separation of energy or length scales affords an appropriate effective field theory description, upon which a perturbative calculation can accurately capture the dynamics of the relevant degrees of freedom.
Exemplars of the success of perturbation theory include the anomalous magnetic moment of the electron and perturbative quantum chromodynamics for jet processes at the Large Hadron Collider~\cite{Aoyama:2019ryr, Aoyama:2020ynm, Boughezal:2022cbl, Maltoni:2022bqs}.

Perturbation theory based on Feynman diagrams suffers, however, from factorial growth in the number of diagrams~\cite{Hurst:1952zh, Bender:1976ni}.
This makes the possibility of performing a realistic calculation in a quantum theory describing everyday interactions quite remote.
Moreover, in absence of precise cancellations, the resulting power series is expected to be divergent~\cite{Dyson:1952tj}.
This phenomenon is present also in classical or tree-level  computations, so that predictions of the classical limit of a perturbative quantum field theory (QFT) do not seem possible, even if we were afforded infinite experimental precision and computational power.
Namely, the divergent asymptotic nature of perturbative expansions in QFT prohibits solving quantum amplitudes to arbitrarily high order in coupling constants.

Nevertheless, some quantum systems are amenable to resummation methods, such as Borel-Laplace resummation, which characterizes and captures the asymptotic nature of the perturbative expansion, as reviewed in Ref.~\cite{Bender1999advanced}.
As the name suggests, the resummation proceeds in two steps, Borel transformation and Laplace transformation.
The former turns a divergent series in a variable $g$ into a series, expressed in a new variable $z$, with nonzero radius of convergence. The resulting series can be analytically continued in the ``Borel plane'', viz.~the complex-$z$ plane. The result of Borel transformation is in turn Laplace-transformed into a \textit{function} valid for any value of $g$.
That is, Borel-Laplace resummation turns an all-order result in perturbation theory from a divergent series into a well-defined function.
This process can fail if the Borel-transformed series has poles on the positive real axis, \textit{i.e.}~in the domain of integration of the Laplace transform (see, \textit{e.g.}, Refs.~\cite{Marino:2012zq, Dorigoni:2014hea, Marino:2021lne} for more details).
Physically, this failure is due to phenomena that lie beyond perturbation theory, such as instantons.
Quantum theories with classically degenerate vacua are a good example of this, as they suffer from the lack of Borel-Laplace summability.
In turn, the location of the poles in the Borel plane gives precise information about instantons, such as their action and relative importance~\cite{Brezin:1977gk, Bogomolny:1977ty, Bogomolny:1980ur, Stone:1977au, Achuthan:1988wh, Liang:1995zq}.

Far from a failure of the program of extracting exact results from perturbation theory, theories that cannot be Borel-Laplace resummed are in the modern perspective studied for the connection they provide between perturbative and nonperturbative data.
Indeed, the specific way in which the resummation procedure can fail encodes nonperturbative effects.
A general theory of such encodings was developed under the term ``resurgence''~\cite{Ecalle1981fonctions1, Ecalle1981fonctions2, Ecalle1981fonctions3}.
Aspects of resurgence in QFT and quantum mechanics (QM) have been widely investigated (see, \textit{e.g.},~\cite{Zinn-Justin:1981qzi, Zinn-Justin:1982hva, Zinn-Justin:1983yiy, Balitsky:1985in, Balitsky:1986qn, Aoyama:1998nt, Aoyama:1997qk, Zinn-Justin:2004vcw, Zinn-Justin:2004qzw, Jentschura:2004jg, Marino:2007te, Marino:2008ya, Unsal:2012zj, Aniceto:2013fka, Dunne:2014bca}).
Applied to high-energy physics, resurgence aims to include nonperturbative effects in the calculation of observables by leveraging its connection to perturbative data.
The latter is typically amenable to a systematic computation in practice.

An asymptotic divergent series in a small parameter $g$ and its resummed counterpart approximately agree for the first $\sim1/g$ orders~\cite{10.1023/A:1006145903624}.
This explains the practical success of perturbation theory and redirects studies of its shortcomings to ``theory laboratories", \textit{i.e.}~questions of internal consistency.
Traditionally, QM has provided an ideal environment for such studies~\cite{Bender:1969si, Bender:1973rz, Loeffel:1969rdm, Zinn-Justin:2004vcw, Dunne:2014bca, Sulejmanpasic:2016fwr, Serone:2017nmd}.
Most of these investigations, however, have focused on the vacuum state, or low-lying excitations of the spectrum.
Here, instead, along the lines of other notable exceptions~\cite{Bachas:1991fd, Brown:1992ay, Bachas:1992dw, Cornwall:1991gx, Cornwall:1992ip, Cornwall:1993rh, Khlebnikov:1994dc, Jaeckel:2018ipq, Jaeckel:2018tdj}, we are interested in highly excited quantum states, \textit{i.e.}~states with energy $E_n$, where $n$ is large.
This is because in the limit $n \to \infty$ and $\hbar \to 0$ we expect all quantum effects to be strongly suppressed, letting us address the aforementioned puzzle of the emergence of classical physics from a quantum world.

The quantum anharmonic oscillator features a single or double well potential, depending on the sign of the quadratic potential term.
For highly energetic quantum states, where the quartic term dominates, the general expectation is that the two sign choices give approximately equal results.
In this sense the classical limits of these two different theories are then expected to be one and the same.
This intuition contrasts with the fact that, at a purely classical level, the theories have different vacuum structure, which should distinguish the two.
In this work we use techniques of resurgence to study the classical limit of the anharmonic oscillator and straighten this apparent contradiction.
Resurgent asymptotics give us the practical bonus of constructing observables from perturbation theory and at the same time removing divergent series from our calculation, which would cast doubts on the validity of any classical limit.
Indeed, this has lead to an \emph{exact} quantization condition for anharmonic oscillators, interpreted as a self-consistency condition on the energy eigenvalues~\cite{Voros1997ExactAQ}.
In particular, we use the method of exact WKB to quantize the anharmonic oscillator.
Based on the Balian-Bloch formalism~\cite{Balian:1974ah}, this method quantizes the theory by considering all classical paths in a \emph{complex} phase space.
The latter drastically contrasts with conventional WKB computations, which only consider classical paths that lie entirely on the real axis.
Paths off the real axis, being classically inaccessible, are then interpreted as purely quantum, typically describing tunneling solutions (instantons).

We anticipate here our findings: even in the classical limit, the amplitudes of the theory contain information on the quantum nature of the system.
We prove this explicitly for the anharmonic oscillator for the case of amplitudes of the type $\braket{n|x|0}$, whose classical limit has been calculated to all orders in Refs.~\cite{Jaeckel:2018ipq,Jaeckel:2018tdj}.
We show that they match exactly the action of the imaginary classical path that enters the quantization condition.
We believe this result -- which we coin the ``quantum imprint rule" -- to hold in any quantum theory, possibly with modifications due to different structure of the vacuum, as we elaborate in the main text.

Our work is arranged as follows.
In \cref{sec:wkb} we briefly review the WKB method and how it is related to the quantization of a classical theory.
Then, in \cref{sec:doublewell}, we apply this to the quantum mechanical anharmonic oscillator featuring a symmetric double well potential.
In particular, we revisit its exact quantization condition by including instantons everywhere in the complex plane.
\Cref{sec:imprintquantum} contains a derivation of the quantum imprint rule, the key result of our work.
We then interpret and discuss the universality and possible extensions of our result to QFT.
Finally, conclusions and prospects for future studies are contained in \cref{sec:conclusions}.

\section{The Exact WKB Method and Quantization of Periods}
\label{sec:wkb}

We review the exact WKB method and its application in QM, with an emphasis on the quantization condition from perturbative and nonperturbative cycles.
We begin, as usual, by constructing wave function solutions to the Schr\"{o}dinger equation as a formal power series in $\hbar$~\cite{10.1103/PhysRev.41.713}, giving asymptotic series expansions of quantized energy level eigenvalues.
Our discussion closely follows Ref.~\cite{Codesido:2017dns}, and a more pedagogical introduction can be found in Ref.~\cite{Marino:2021lne}.

\subsection{The WKB approach}
\label{subsec:WKBapproach}

The general idea behind the WKB method is that the quantum action, controlled by the expansion parameter $\hbar$, is small compared to the classical action associated to the particle's motion.
That is, $\hbar$ is small compared to the classical phase-space volume occupied by the particle, and as such, the WKB ansatz is a semiclassical expansion.
Quantum mechanically, this corresponds to particle states at large quantum numbers, occupying many quanta of the available phase space.

In QM, the dynamics of a particle moving in a potential $V(x)$ are governed by the Schr\"{o}dinger equation,
\begin{equation}
	\hbar^2 \psi^{\prime \prime} (x) + p^2(x) \psi (x) = 0 \, ,
\label{eq:schroedinger}
\end{equation}
where $\psi(x)$ denotes the particle's wave function, the prime denotes a derivative with respect to $x$, the mass has been set to unity, and $p$ is the momentum associated to the energy $E$ of the particle,
\begin{equation}
	p(x) = \sqrt{2 \left(E - V(x) \right)} \, .
\end{equation}
Using the WKB approach, we write $\psi(x)$ as
\begin{equation}
	\psi(x) = \exp \left( \frac{i}{\hbar} \int^x \md y \, Q(y) \right) \, ,
\label{eq:wkbansatz}
\end{equation}
where $Q(x)$ is an unknown function.
The WKB ansatz transforms \cref{eq:schroedinger} into a Riccati equation on $Q(x)$,
\begin{equation}
	Q^2 - i \hbar Q^{\prime} = p^2(x) = 2 \left(E - V(x) \right) \, .
\label{eq:riccati}
\end{equation}
At this level, we are simply rewriting the Schr\"{o}dinger equation, but the chosen form for $\psi(x)$ is particularly apt for a semiclassical expansion.
In the limit in which $\hbar$ is small, the wave function becomes highly oscillatory, \textit{i.e.}~its de~Broglie wavelength becomes unobservably small.
Since analytic solutions to \cref{eq:riccati} for general $V(x)$ are lacking, we expand $Q(x)$ as a power series in $\hbar$,
\begin{equation}
	Q(x) = \sum_{k=0}^{\infty} Q_k(x) \hbar^k \, ,
\label{eq:WKBSeries}
\end{equation}
which, according to \cref{eq:riccati}, dictates a recursion relation of the series coefficients of $Q(x)$,
\begin{equation}
	Q_{k+1}(x) = \frac{1}{2 Q_0(x)} \left(i Q_k^{\prime}(x) - \sum_{i=1}^{k} Q_i(x) Q_{k+1-i}(x) \right) \, ,
\label{eq:WKBDifferential}
\end{equation}
with $Q_0(x) = \pm p(x)$.
Here, the two sign choices generate two independent solutions of \cref{eq:riccati}.
This recursion relation enables us to reconstruct the wave function to arbitrary order in $\hbar$, in principle.

We can now notice that not all terms of the series expansion~\eqref{eq:WKBSeries} carry relevant information.
If we split the series expansion into contributions of even and odd powers of $\hbar$,
\begin{equation}
	Q(x) = \sum_{k=0}^\infty Q_{2k}(x) \hbar^{2k} + \sum_{k=0}^\infty Q_{2k+1}(x) \hbar^{2k+1} \equiv P(x) + Q_{\mathrm{odd}}(x) \, ,
\end{equation}
we find that the odd contributions are a total derivative,
\begin{equation}
	Q_{\mathrm{odd}}(x) = \frac{i \hbar}{2} \frac{P^{\prime}(x)}{P(x)} = \frac{i \hbar}{2} \frac{\md}{\md x} \log P(x) \, .
\end{equation}
Therefore, we can drop the odd contribution and write
\begin{equation}
	\psi^{\pm} (x) = \frac{1}{\sqrt{P(x)}} \exp \left(\pm \frac{i}{\hbar} \int^x \md y \, P(y) \right) \, ,
\label{eq:WKBWaveFunction}
\end{equation}
where the sign corresponds to the sign choice made on $Q_0$.
Strictly speaking, we have only rewritten a solution of the Schr\"{o}dinger equation in terms of a formal power series expansion in the quantum action.
We have not yet made a step towards describing true \emph{quantum} solutions of the system, in the sense that the eigenfunctions of the Hamiltonian have to span a Hilbert space, which is only true if the particle's energy is quantized according to $\hbar$.
In the WKB framework, the necessary condition is implemented via the quantization of so-called quantum periods, which we discuss next.

\subsection{Connection problems, quantum periods, and quantization}
\label{subsec:quantization}

The linearity of the Schr\"odinger equation, together with the two possible sign choices in \cref{eq:WKBWaveFunction}, implies that the most general form of the wave function is
\begin{equation}
 	\psi(x) = a^+\psi^+(x) + a^-\psi^-(x) \, .
\label{eq:WKBWaveFunctionFull}
\end{equation}
Asymptotic boundary conditions then determine the coefficients $a^{\pm}$, since the sign of the imaginary part of $p(x)$ dictates whether a solution grows without bounds (``dominant") or vanishes (``recessive") at infinity.
Due the normalizability of the wave function, only a recessive solution is viable at infinity and the behavior of its coefficient can be uniquely established.
That is, asymptotic boundary conditions at infinity only specify the recessive solution, leaving the dominant contribution unspecified~\cite{Voros1983}.
The recessive solution is then analytically continued from infinity.
However, if a point is encountered where $p(x)$ vanishes along this process, either of the solutions changes discontinuously, since the integration region of the exponent in \cref{eq:WKBWaveFunction} now includes a branch point of the integrand.
Beyond the branch point, a different linear combination of $\psi(x)^+$ and $\psi(x)^-$ is employed, such that requiring continuity of the wave function casts the WKB method into a ``connection problem", \textit{i.e.}~the problem of finding the coefficients $a^{\pm}$ in one region given a set of known coefficients in another region.
A complete solution of this connection problem requires analytical control over exponentially suppressed quantities around the branch points of $p(x)$~\cite{Froman1965jwkb}.
Indeed, if such a solution can be found, it is expressed in terms of a connection formula.

The points where $p(x)$ vanishes are called ``turning points", because in a classical theory they are the points where kinetic energy vanishes and the particle reverses its motion.
Due to conservation of energy, they are given by
\begin{equation}
	p(x)^2 = E - V(x) = 0 \, .
\end{equation}
A region where $p(x) > 0 $ is then classically allowed, while a region where $p(x) < 0$ is classically forbidden.
As a simple example, let us consider a confining potential $V(x)$ that features exactly two turning points, $x_-$ and $x_+$.
In the quantum theory, we expect the wave function of the particle to be highly oscillating in the classically allowed region inside the well and exponentially decaying outside of it.
This behavior has to be mimicked by the WKB ansatz~\eqref{eq:WKBWaveFunction} and, as explained before, the two different behaviors have to match at the turning points of the potential.
Crucially, in a scenario involving only two turning points, solving this connection problem leads to the quantization condition (see, \textit{e.g.}, Ref.~\cite{Marino:2021lne})
\begin{equation}
	\int_{x_-}^{x_+} \md x \, P(x) = \pi \hbar \left( n + \frac{1}{2} \right) \, ,
\label{eq:ExactQuantization2TP}
\end{equation}
where $n$ is an integer.
We stress that this expression is \emph{exact}.
At the same time, the left hand side is given as a formal power series in $\hbar$.
An approximation of the latter by its leading order term leads to the well-known Bohr-Sommerfeld quantization,
\begin{equation}
	\int_{x_-}^{x_+} \md x \, p(x) = \pi \hbar \left( n + \frac{1}{2} \right) \, .
\label{eq:BohrSommerfeld}
\end{equation}
This can be thought of as a quantization of the phase space volume that is occupied by the particle of energy $E$, as given by the left hand side of this relation.
The solutions to this equation finally give the quantized energy levels, $E = E_n$.
In this sense, the exact quantization condition~\eqref{eq:ExactQuantization2TP} incorporates higher order corrections to the Bohr-Sommerfeld formula.

In more geometric terms, the integral between the two turning points can be written as an integral along a contour $\mathcal{A}$ in the complex plane, which encircles the allowed classical motion of the particle at the corresponding energy,
\begin{equation}
	\int_{x_-}^{x_+} \md x \, P(x) = \frac{1}{2} \oint_{\mathcal{A}} \md x \, P(x) \, .
\label{eq:QuantumPeriod}
\end{equation}
Closing the contour in the complex plane regularizes the integral on the left hand side, which otherwise would suffer from divergences at the turning points, because these are branch points of the integrands (see, \textit{e.g.}, Ref.~\cite{Marino:2021lne}).
The right hand side of \cref{eq:QuantumPeriod} is typically referred to as a ``quantum" or ``WKB period."\footnote{The name is chosen in analogy to integrals that appear in the calculation of planetary orbits.}
Using the formal series expansion of $P(x)$ in powers of $\hbar$, the exact quantization condition can be written as
\begin{equation}
	 \oint_{\mathcal{A}} \md x \, P(x) = \sum_{k=0}^{\infty} \vol_k (E) \hbar^{2k} = 2 \pi \hbar \left( n + \frac{1}{2} \right) \, ,
\label{eq:ExactQuantization2TPSeries}
\end{equation}
where we have defined the quantum coefficients of the occupied phase space volume,
\begin{equation}
	\vol_k (E) = \oint_{\mathcal{A}} \md x \, Q_{2k}(x) \, .
\label{eq:volnE}
\end{equation}
These volumes substantially enter connection formulas, as we have already seen for the case of a potential with two turning points.
Implementing them in practice, however, is not straightforward.
Generically, the series expansion in \cref{eq:ExactQuantization2TPSeries} is factorially divergent, $\vol_k(E) \sim (2k)!$.
The situation is even more dire, as it is typically not even Borel-Laplace resummable.
Therefore, we have to consider exponentially suppressed corrections to the formal power series responsible for lifting the ambiguities that lead to the failure of Borel-Laplace summability.

The problem is even more pronounced in QM systems involving more than two classical turning points.
The pairwise matching of turning points to build quantum periods must also be tessellated to cover the entire Borel plane~\cite{Voros1997ExactAQ}, since a consistent connection formula needs to take into account all possible periods~\cite{10.1103/PhysRevLett.41.1141}.
As a blessing in disguise, the inclusion of these periods in turn corresponds to the aforementioned exponentially small corrections that enable unambiguous resummation.
For a concise introduction to the theory of asymptotic series and Borel-Laplace resummation, we refer the reader to \cref{app:BorelLaplace}, while a more detailed introductory account can be found in Refs.~\cite{Marino:2012zq, Dorigoni:2014hea, Marino:2021lne}.

\section{The Exact Quantization Condition in the Symmetric Double Well}
\label{sec:doublewell}

Applying the exact WKB method in practice requires analyzing situations with multiple turning points, which can be associated with perturbative or nonperturbative phenomena in the quantum theory.
We will study these details in the context of the double well potential following Ref.~\cite{Marino:2021lne}.

\subsection{Classical turning points and their structure}
\label{ssec:TurnPoint}

The symmetric double well is a prime example from QM where nonperturbative effects play a crucial role, since the instanton solutions breaks the classical denegeracy of equivalent vacua.
Said degeneracy is conserved in the energy levels derived from perturbation theory and only broken nonperturbatively  (see, \textit{e.g.}, Refs.~\cite{Zinn-Justin:2004vcw, Zinn-Justin:2004qzw}).
We remark that QM models with smooth potentials cannot undergo spontaneous symmetry breaking, because instantons explicitly break the classical symmetry enjoyed by degenerate vacua.
Thus, their effect in understanding the physics of QM systems is irreducible.
The exact WKB method systematically accounts for instanton effects and results in a recipe to unambiguously calculate observables whose series expansion is otherwise not Borel-Laplace resummable.

We parameterize the symmetric double well potential as
\begin{equation}
	V_{\mathrm{DW}}(x) = \frac{1}{32g^2} \left( 1 - 4g^2 x^2\right)^2 \, .
\label{eq:DWPotential}
\end{equation}
The classical vacua are located at $v_{\pm} = \pm 1 / (2g)$, with an energy barrier of height $\Lambda = 1/(32g^2)$ separating them.
The potential has four classical turning points.
Due to the $\mathbb{Z}_2$ symmetry of the theory, the turning points come in pairs and, for given $E$, are
\begin{equation}
	a^2 = \frac{1 + 4 \sqrt{2Eg^2}}{4g^2} \, , \enspace b^2 = \frac{1 - 4 \sqrt{2Eg^2}}{4g^2} \, .
\label{eq:DWTurningPoints}
\end{equation}
Note that $b$ becomes imaginary if the particle's energy exceeds the potential barrier, $E > \Lambda$.
Thus, the turning point structure of the double well characterizes classically allowed and forbidden regions.

Intuitively, a particle with energy less than $\Lambda$ will oscillate around one of the two minima between the turning points $x \in (-a, -b)$ or $x \in (b, a)$.
Quantum mechanically, of course, the particle can tunnel through the classically forbidden region $(-b,b)$ to the other well.
The classically allowed (forbidden) region is associated to the WKB contour $\mathcal{A}$ ($\mathcal{B}$) from \cref{eq:QuantumPeriod}, lying on the real axis and encircling the corresponding turning points.

We now consider $E > \Lambda$, such that the particle can cross over the barrier to the second well.
The classically allowed region, characterized by the cycle $\mathcal{A}$, now extends from $x = -a$ to $x = a$, while the formerly forbidden region with $x \in (-b, b)$ has moved to the imaginary axis in the complex plane.
In this case, the contour $\mathcal{B}$ does not describe below-the-barrier tunneling anymore, instead corresponding to above-the-barrier reflection.
Nevertheless, it contains crucial physical information such as level splitting.

\subsection{Quantum periods of the double well}

As we have outlined in \cref{sec:wkb}, wave function continuity everywhere in the complex plane is essential for quantization, and it can be expressed in the form of a connection formula.
In terms of resurgent language, the Voros-Silverstone connection formulae~\cite{Voros1981SpectreDL, 10.1103/PhysRevLett.55.2523} provide the solution to this problem as well as the conceptual basis for the complex WKB framework.
For the symmetric double well, the Voros-Silverstone connection formula is translated into the exact quantization condition~\cite{Voros1983} 
\begin{equation}
	1 + \me^{\pm 2 \pi i \nu} = \pm \epsilon i f(\nu) \, ,
\label{eq:DWQuantization}
\end{equation}
where
\begin{equation}
	\nu = \frac{1}{2\pi\hbar} \oint_{\mathcal{A}} \md x \, P(x) \, , \enspace f(\nu) = \exp \left( \frac{i}{2\hbar} \oint_{\mathcal{B}} \md x \, P(x) \right) \, .
\label{eq:nu_fnu}
\end{equation}
Both sides of \cref{eq:DWQuantization} depend on the energy of the particle via the definitions of the contours $\mathcal{A}$ and $\mathcal{B}$ following \cref{eq:QuantumPeriod,eq:DWTurningPoints} and prescribe the exact solutions and energies of the quantum theory.

\subsection{Perturbative quantum periods}
\label{ssec:PertQuant}

If we momentarily neglect the nonperturbative cycle and set $f(\nu)=0$, the quantization condition simplifies to the well-known Bohr-Sommerfeld quantization presented in \cref{sec:wkb},
\begin{equation}
	\nu = n + \frac{1}{2} \, ,
\end{equation}
where $n$ is a positive integer, such that the perturbative cycle satisfies
\begin{equation}
	\oint_{\mathcal{A}} \md x \, P(x) = 2 \pi \hbar \left( n + \frac{1}{2} \right) \, .
\label{eq:QuantizationP}
\end{equation}
To obtain energy eigenvalue solutions that satisfy \cref{eq:QuantizationP}, we adopt the systematic approach of~\cite{Fischbach:2018yiu} and first expand the left hand side in $\hbar$ as $\oint_{\mathcal{A}} \md x \, P(x) = \sum_{k=0}^{\infty} \vol_k(E) \hbar^{2k}$.

Geometrically, for the symmetric double well, the kinetic energy of the particle $p^2(x)$ describes an elliptic curve of genus one in the complex plane of $x$.
This implies that the different quantum corrections to the phase space volume, $\vol_k(E)$, can be generated from a differential operator of at most second order acting on $\vol_0(E)$~\cite{Basar:2017hpr, Fischbach:2018yiu},
\begin{equation}
	\vol_k (E) = \left( f_k^{(0)} + f_k^{(1)} \frac{\md}{\md E} + f_k^{(2)} \frac{\md^2}{\md E^2} \right) \mathrm{vol}_0 (E) \, ,
\label{eq:volktovol0}
\end{equation}
where the coefficients are functions of the energy, $f_k^{(i)}(E)$.

Recalling that $\vol_k(E) = \oint_{\mathcal{A}} \md x \, Q_{2k} (x)$, and $Q_{2k} (x)$ are known via the recursion relation~\eqref{eq:WKBDifferential}, we can adapt the recursion relation to identify the $f_k^{(i)}$ functions using the ansatz~\cite{Fischbach:2018yiu}
\begin{equation}
	\sum_{i=0}^2 f_k^{(i)} \frac{\md^i}{\md E^i} Q_0(x) = Q_{2k}(x) + \sum_{j=0}^{j_{\mathrm{max}}} \frac{\partial}{\partial x} \left( \alpha_j (E) \frac{x^j}{p^{3k-3+2r}(x)} \right) \, ,
\label{eq:PFAnsatz}
\end{equation}
for suitable coefficients $\alpha_j (E)$ and integer $r$, and $j_{\mathrm{max}}$ large enough.
More precisely, as any choice of $r$ and $j_{\mathrm{max}}$ fixes a set of solutions $\alpha_j$ and $f_k^{(i)}$, the former has to be chosen large enough such that the recursion ansatz~\eqref{eq:PFAnsatz} closes.
This recasting of the recursion relation from the Riccati equation into geometric language is guaranteed by arguments based on cohomology~\cite{Fischbach:2018yiu, Kreshchuk:2018qpf}.

For illustration, we present the first coefficient functions $f_1^{(i)}(E)$ for the symmetric double well.
As a starting point, the leading-order volume reads~\cite{Gradshteyn2007Integrals}
\begin{equation}
	\mathrm{vol}_0(E) = \oint_{\mathcal{A}} p(x) = 2 \int_{b}^{a} p(x) = \frac{2}{3} g a \left[ \left(a^2 + b^2\right) E \left(1 - \frac{b^2}{a^2} \right) - 2b^2 K \left(1 - \frac{b^2}{a^2}\right) \right] \, ,
\end{equation}
where $K$ and $E$ denote the complete elliptic integral of the first and second kind, respectively, and the turning points $a$ and $b$ are given by \cref{eq:DWTurningPoints}.
The ansatz~\eqref{eq:PFAnsatz} then gives a prescription to compute the first quantum correction to the classical period,
\begin{equation}
	\mathrm{vol}_1(E) = \left(- \frac{g^2 \left(48 E g^2 - 1\right)}{4E\left(32 E g^2 - 1\right)} + \frac{g^2}{2} \frac{\md}{\md E} \right) \mathrm{vol}_0(E) \, .
\label{eq:PFOperatorExplicit}
\end{equation}
We remark that this operator is not unique, as we could have replaced the constant term $f_1^{(0)}(E)$ by a second-derivative term $f_1^{(2)}(E)$.

Including the first quantum correction $\vol_1(E)$, we obtain the series expansion of \cref{eq:QuantizationP} in terms of the energy and coupling,
\begin{equation}
	\oint_{\mathcal{A}} \md x \, P(x) = 2 \pi E + \left( 6 \pi E^2 + \frac{1}{2} \pi \hbar^2 \right) g^2 + \left( 70 \pi E^3 + \frac{25}{2} \pi E \hbar^2 \right) g^4 + \ldots \, .
\label{eq:QuantizationPSeries}
\end{equation}
After including more orders of the quantum corrections to the perturbative cycle, we invert the series expansion~\eqref{eq:QuantizationPSeries} and obtain the standard result of Rayleigh-Schr\"odinger perturbation theory for the energy levels,
\begin{equation}
	E(\nu) = \hbar \nu - \left( 3 \nu^2 + \frac{1}{4} \right) \hbar^2 g^2 - \left( 17 \nu^3 + \frac{19}{4} \nu \right) \hbar^3 g^4 + \mathcal{O} \left(\hbar^4 g^6\right) \, ,
\label{eq:DWEnergyP}
\end{equation}
as presented in the seminal work by Bender and Wu~\cite{Bender:1969si, Bender:1973rz}, whose methodology has been implemented in Ref.~\cite{Sulejmanpasic:2016fwr}.
We remark that the non-alternating signs of the leading-$\nu$ coefficients indicate that the series is not Borel-Laplace resummable, ultimately due to having neglected instanton contributions.

Before we close this discussion, we note that the leading-order terms of the form $\nu^j \hbar^j g^{2(j-1)}$ in \cref{eq:DWEnergyP} are generated by the leading-order period $\vol_0(E)$ upon series inversion.
Indeed, this reflects that textbook WKB is a semiclassical expansion, suitable for quantum states at large $\nu$.
However, terms subleading in $\nu$ at each order of the expansion can be obtained from higher order corrections to the quantum period, $\vol_{k\geq1}(E)$.
In this way, the all-order (perturbative) result obtained via WKB reproduces Rayleigh-Schr\"odinger perturbation theory for arbitrary quantum numbers $\nu$.
What is still lacking is an account of the nonperturbative corrections that arise when $f(\nu) \neq 0$.

\subsection{Nonperturbative quantum periods}

To systematically account for the nonperturbative quantum effects, we now have to include the function $f(\nu)$ in \cref{eq:DWQuantization}.
We first compute its exponent, \textit{i.e.}~the nonperturbative quantum period associated to the cycle $\mathcal{B}$.
Therefore, we expand
\begin{equation}
	-i \oint_{\mathcal{B}} \md x \, P(x) = \sum_{k=0}^{\infty} \vol_k^{\mathrm{np}} (E) \hbar^{2k} \, ,
\label{eq:DWVolumeExpansionNP}
\end{equation}
where we have defined the coefficients of the nonperturbative quantum period as
\begin{equation}
	\vol_k^{\mathrm{np}} (E) = -i \oint_{\mathcal{B}} \md x \, Q_{2k} (x) \, .
\label{eq:volnNPE}
\end{equation}
These coefficients can be computed in the same manner as the coefficients of the perturbative quantum period presented previously.
Starting at leading order we get~\cite{Gradshteyn2007Integrals}
\begin{equation}
	\vol_0^{\mathrm{np}}(E) = -i\oint_{\mathcal{B}} p(x) = -2i \int_{-b}^{b} p(x) = \frac{4}{3} g a \left[ \left(a^2 + b^2\right) E \left(\frac{b^2}{a^2} \right) - \left(a^2 - b^2\right) K \left(\frac{b^2}{a^2}\right) \right] \, .
\end{equation}
This is indeed explicitly related to the corresponding perturbative cycle, $\vol_0(E)$, as prescribed by the geometry of genus-one curves (see Ref.~\cite{Basar:2017hpr} and references therein).
Moreover, the differential operator that generates $\vol_k^{\mathrm{np}} (E)$ exactly coincides with the one that generates $\vol_k(E)$, essentially because they are expressed as closed contour integrals in the complex plane and their differential operators may be taken outside the integral.
For instance, the first quantum correction to the nonperturbative period is, similar to \cref{eq:PFOperatorExplicit},
\begin{equation}
	\vol_1^{\mathrm{np}} (E) = \left(- \frac{g^2 \left(48 E g^2 - 1\right)}{4E\left(32 E g^2 - 1\right)} + \frac{g^2}{2} \frac{\md}{\md E} \right) \vol_0^{\mathrm{np}} (E) \, ,
\end{equation}
with all higher orders working similarly.

Next, closely following~\cite{Marino:2021lne}, we plug the perturbative series expression for the energy $E(\nu)$, shown in \cref{eq:DWEnergyP}, into the nonperturbative quantum period via the turning points.
We reproduce a series expansion of the form
\begin{equation}
    \begin{split}
        - \frac{i}{\hbar} \oint_{\mathcal{B}} \md x \, P(x) = & \frac{1}{3 g^2 \hbar} \\
        & + \frac{1}{\hbar} \left(2 \left( \log \left( \frac{g^2 \nu \hbar}{2} \right) - 1 \right) \nu \hbar + 17 g^2 \nu^2 \hbar^2 + 125 g^4 \nu^3 \hbar^3 + \frac{17815}{12} g^6 \nu^4 \hbar^4 + \ldots \right) \\
        & + \hbar \left( - \frac{1}{12 \nu \hbar} + \frac{19}{12} g^2 + \ldots \right) \\
        & + \hbar^3 \left( \frac{7}{1440 \nu^3 \hbar^3} + \frac{22709}{576} g^6  + \ldots\right) + \mathcal{O} \left(\hbar^5\right) \, .
    \end{split}
\label{eq:DWQuantumPeriodNPSeries}
\end{equation}
The organization of terms according to $\hbar$ allows us to identify the different contributions from each quantum volume correction.
For example, in the right hand side of \cref{eq:DWQuantumPeriodNPSeries}, terms proportional to $1 / \hbar$ (with powers of $\nu \hbar$ fixed) originate from the leading order period, $\vol_0^{\mathrm{np}}(E)$, and in general the terms proportional to $\hbar^{2k-1}$ correspond to $\vol_{k}^{\mathrm{np}}(E)$.

Recalling that $f(\nu) = \exp [ i \oint_{\mathcal{B}} \md x \, P(x) / (2 \hbar)]$, we see that the leading contribution of $f(\nu)$ to the quantization condition is an exponentially small shift of $\nu$ away from half-integers,
\begin{equation}
	f(\nu) \propto \exp \left( - \frac{1}{6g^2\hbar} + \ldots \right) \, .
\end{equation}
This contribution can be recognized as the instanton action, confirming that $f(\nu$) encodes the quantum effects of tunneling between the wells of the potential.
In fact, the inclusion of nonperturbative effects through $f(\nu$) lifts the vacuum degeneracy between the energy levels $E_{\nu}^{\pm}$ of opposite parity (see, \textit{e.g.}, Ref.~\cite{Marino:2021lne}), since
\begin{equation}
	E_{\nu}^+ - E_{\nu}^- = - \frac{1}{\pi} f(\nu) \frac{\partial E(\nu)}{\partial \nu} + \ldots
\label{eq:EnergySplitting}
\end{equation}
is an explicit instanton-induced symmetry breaking of the classical $\mathbb{Z}_2$ symmetry.

Besides the explicit calculable correction to energy levels and the resolution of the $\mathbb{Z}_2$ parity breaking and Borel-Laplace summability questions, the nonzero nature of $f(\nu)$ is evident in all physical quantities, even when evaluating amplitudes corresponding to a classical limit of the theory.
In the next section, we justify this claim, which is the main result of this work.

\section{Quantum Imprints at Large Quantum Number}
\label{sec:imprintquantum}

In this section, we demonstrate a novel correspondence between the tunneling action in the anharmonic oscillator and the amplitude of multiparticle production in scalar QFT.
We are particularly focused on the expression from \cref{eq:DWQuantumPeriodNPSeries}, which admits two different scaling regimes as a formal series expansion in $\hbar$.

The common approach~\cite{Marino:2021lne} is to consider $\nu$ fixed, and send $\hbar \to 0$ to approach a classical regime.
For fixed $\nu$ being small, however, it is doubtful that semiclassical behavior is obtained given that energy eigenstates are nonclassical.
Na\"{i}vely, the semiclassical limit is obtained for large $\nu$, where the action is large with respect to $\hbar$.
As a result, simultaneously considering $\nu \to \infty$ and $\hbar \to 0$ suppresses quantum behavior.
Moreover, the correct states to consider in the semiclassical limit are coherent states, \textit{i.e.}~states with minimal uncertainty.
In particular, coherent states, by virtue of saturating the Heisenberg uncertainty principle, become classical in the $\hbar \to 0$ limit.
On the other hand, coherent states do not have a well-defined occupation number $\nu$, but instead follow a Poisson distribution centered around a given $\nu$ with a width of $\sqrt{\nu}$.
At large $\nu$, the width divided by the mean occupation number approaches zero, and so does the distinction between $\ket{\nu}$ and $\ket{\nu + 1}$.
Hence, in the $\nu \to \infty$ and $\hbar \to 0$ double scaling limit, using coherent or energy eigenstates gives the same amplitude.

We are thus motivated to define the 't~Hooft-like coupling parameter
\begin{equation}
	\lambda = \hbar \nu \, ,
\label{eq:lambdadef}
\end{equation}
which stays fixed in the double scaling limit $\nu \to \infty$ and $\hbar \to 0$.
In this double scaling limit, terms proportional to $1/\hbar$ in \cref{eq:DWQuantumPeriodNPSeries} are the only relevant contribution to the nonperturbative quantum period, and its exponent takes the form
\begin{equation}
    \begin{split}
    \left. 
    - \frac{i}{\hbar} \oint_{\mathcal{B}} \md x \, P(x)
    \right\rvert_{\frac{1}{\hbar}} &\simeq 
    \frac{1}{3 g^2 \hbar} 
    \\
    &+ \frac{\lambda}{\hbar} \left[
    2 \left( \log \left( \frac{g^2 \lambda}{2} \right) - 1 \right) + 17 g^2 \lambda + 125 g^4 \lambda^2 + \frac{17815}{12} g^6 \lambda^3 + \mathcal{O} \left( g^8 \lambda^4 \right)
    \right]    \, .
    \end{split}
\label{eq:DWQuantumPeriodHolyGrail}
\end{equation}
We remark that \cref{eq:DWQuantumPeriodHolyGrail}, as well as \cref{eq:DWEnergyP} depend on the perturbation coupling $g^2$ via powers of $g^2 \hbar$.
This affords a possible rescaling in the anharmonic oscillator~\cite{Simon:1970mc} to absorb $\hbar$ into $g^2$.
Here, we do not perform this substitution because we are interested in a semiclassical limit, which is made more explicit by keeping $g^2$ fixed and sending $\hbar \to 0$.

\subsection{The quantum imprint rule}
\label{subsec:quantumimprintrule}

We now demonstrate that the double scaling limit shown in \cref{eq:DWQuantumPeriodHolyGrail} is remarkably related to the multiparticle production amplitudes at very high energies in QFT.
In particular, in $\phi^4$ theory, the amplitude of a highly-excited scalar decaying to a high multiplicity final state of $n$ on-shell scalars, $\phi^{\ast} \to n \phi$, provides an example of nonperturbative UV dynamics in QFTs.
As the effective expansion parameter $g^2 n$ is no longer small, the process becomes nonperturbative in the large $n$ regime given $n$ is proportional to the energy~\cite{Cornwall:1990hh, Goldberg:1990qk, Brown:1992ay, Voloshin:1992mz, Argyres:1992np, Voloshin:1992nu, Smith:1992rq, Gorsky:1993ix, Argyres:1993wz, Libanov:1994ug, Son:1995wz, Khoze:2014zha, Khoze:2017ifq}.
In this scenario, it is believed that the multiparticle amplitude is of exponential form, which could either be decaying or even growing as $n$ becomes large~\cite{Voloshin:1992nu, Khlebnikov:1992af, Libanov:1994ug, Libanov:1995gh, Bezrukov:1995qh, Schenk:2021yea, Khoze:2022fbf} (see also Ref.~\cite{Khoze:2018mey} for a review).
A quantum mechanical analogue to this amplitude is given by the matrix element $\braket{\nu | x | 0}$ of the anharmonic oscillator with a quartic coupling.
Such matrix elements were shown to be generically of exponential form inversely proportional to the anharmonic coupling $g^2$~\cite{Jaeckel:2018ipq, Jaeckel:2018tdj},
\begin{equation}
	\braket{\nu | x | 0} = \exp \left( \frac{F}{g^2} \right) \, ,
\end{equation}
where the ``holy grail function" $F$ is a function of the coupling $g^2$ and the energy level $\nu$.
For the double well potential~\eqref{eq:DWPotential} at large $\nu$, the function $F$ reads~\cite{Jaeckel:2018tdj},\footnote{Here, we adopt a different normalization for the mass and coupling in the Schr\"odinger equation compared to Ref.~\cite{Jaeckel:2018tdj}, and we explicitly keep track of factors of $\hbar$.}
\begin{equation}
	\frac{F}{g^2} = \frac{1}{2} \left( \log \left( \frac{g^2 \lambda}{2} \right) - 1 \right) \frac{\lambda}{\hbar} + \frac{17}{4} \frac{g^2 \lambda^2}{\hbar} + \frac{125}{4} \frac{g^4 \lambda^3}{\hbar} + \frac{17815}{48} \frac{g^6 \lambda^4}{\hbar} + \ldots + \mathcal{O}\left(\frac{1}{\nu}\right) \, .
\label{eq:HolyGrail}
\end{equation}
Remarkably, up to an overall numerical factor of four, all but the first term $1/(3g^2 \hbar)$ of the nonperturbative quantum period in \cref{eq:DWQuantumPeriodHolyGrail} can be read off from $F$.

Therefore, in the double scaling limit with fixed 't~Hooft coupling $\lambda$, we find the curious relation
\begin{equation}
	\lim_{\substack{\nu \to \infty \\ \hbar \to 0}} \braket{\nu | x | 0}^2 f (\nu) = \exp \left(-\frac{1}{6g^2 \hbar}\right) \, ,
\label{eq:NPHolyGrail}
\end{equation}
for any fixed $\lambda$.
Here, the right hand side of~\cref{eq:NPHolyGrail} corresponds to the instanton action associated to the tunneling of the quantum theory and is, in particular, independent of the quantum number $\nu$.
In other words, we see that in the limit of large $\nu$, the transition amplitude from the vacuum to a highly excited state, driven by the operator $x$, multiplied by the nonperturbative quantum period is an exact one-instanton amplitude.
We have verified this explicitly for the first 15 terms of both \cref{eq:DWQuantumPeriodHolyGrail} and \cref{eq:HolyGrail}.
In the following, we refer to~\cref{eq:NPHolyGrail}, the key result of our work, as the ``quantum imprint rule."

\subsection{Discussion of the quantum imprint rule}
\label{subsec:discussion}

To fully appreciate the physical insights provided by \cref{eq:NPHolyGrail}, we first revisit the exact quantization condition~\eqref{eq:DWQuantization} for the symmetric double well potential.
The right hand side, $f(\nu)$, measures the tunneling through the barrier that separates the wells.
It solely corresponds to the quantum contribution to the wave function continuity, or in other words, setting $f(\nu) = 0$ resembles a classical theory.
Thus, \cref{eq:DWQuantization} can be understood as a quantization condition for standing waves where the boundary conditions specify the spectrum of modes. 
This is equivalent to the perturbative quantization of the latter via \cref{eq:QuantizationP}, which would, for instance, ignore the splitting of the energy levels through instanton effects.

And yet, only a nonvanishing $f(\nu)$ can provide information about the intrinsic quantum nature of the theory, which cannot be absent according to \cref{eq:NPHolyGrail}.
That is, for physical observables such as the transition amplitude $\braket{\nu | x | 0}$ in the semiclassical regime $\nu \to \infty$ and $\hbar \to 0$, the leading instanton contribution on the right hand side of \cref{eq:NPHolyGrail} prevails: the amplitude is asymptotically given by the nonperturbative cycle, \textit{i.e.}~a solely quantum contribution to the theory, multiplied by the one-instanton amplitude,
\begin{equation}
    \braket{\nu | x | 0}^2 \sim \exp \left( - \frac{1}{6g^2\hbar} \right) \exp \left( - \frac{i}{2\hbar} \oint_{\mathcal{B}} \md x \, P(x) \right) \quad \text{for} \quad \nu \to \infty \, , \, \hbar \to 0 \ .
\end{equation}
This leading deviation from classical behavior is always present for all finite coupling values, and formally ``gaps" amplitudes in the quantum theory from amplitudes in a classical analogue.

Moreover, the quantum imprint rule is an exact result in the coupling expansion by construction, since it involves both the all-order perturbative result of $\braket{\nu | x | 0}$ as well as the nonperturbative quantum period entering the function $f(\nu)$, precluding any further quantum corrections.
We carefully note, however, that this only holds in the limits $\nu \to \infty$ and $\hbar \to 0$.

Indeed, both $\braket{\nu | x | 0}$ and $f(\nu)$ receive corrections in powers of $1 / \nu$ (or equivalently in $\hbar / \lambda$), that do not obey the relationship observed in~\cref{eq:NPHolyGrail}.
Explicitly, the first correction to the nonperturbative quantum period in the double scaling limit involves terms proportional to $\hbar$,
\begin{equation}
	\left. - \frac{i}{\hbar} \oint_{\mathcal{B}} \md x \, P(x) 
 \right\rvert_{\hbar} \simeq 
 g^2 \hbar \left( 
 \frac{19}{12} + \frac{153}{4} g^2 \lambda + \frac{23405}{24} g^4 \lambda^2 + \mathcal{O}\left(g^6 \lambda^3\right)
 \right) \, .
\label{eq:CorrNPPeriod}
\end{equation}
Similarly, the holy grail function $F$ was shown to enter the multiparticle amplitude $\braket{\nu | x | 0}$ in the form~\cite{Jaeckel:2018ipq, Jaeckel:2018tdj}
\begin{equation}
	\braket{\nu | x | 0} = \exp \left( \frac{F}{g^2} \right) = \exp \left[ \frac{1}{g^2} \left( F_0 + \frac{F_1}{\nu} + \frac{F_2}{\nu^2} + \ldots \right) \right] \, .
\end{equation}
Upon reintroducing factors of $\hbar$, the first quantum correction to the exponent in our notation reads
\begin{equation}
	\frac{1}{g^2} \frac{F_1}{\nu} = \frac{5}{16} g^2 \lambda + \frac{99}{128} g^4 \lambda^2 + \frac{4279}{1536} g^6 \lambda^3 + \ldots \, .
\label{eq:CorrF1}
\end{equation}
Recalling that $\lambda = \hbar \nu$, we find that the correction to $f(\nu)$ given in \cref{eq:CorrNPPeriod} is $1/\nu$-suppressed with respect to the quantum correction~\eqref{eq:CorrF1}.
Therefore, the two expressions do not match, as they come with different powers of $1/\nu$.
Moreover, this pattern continues beyond the leading quantum $1/\nu$ (or $\hbar$) corrections to the quantum imprint rule.
In particular, the holy grail function $F$ involves terms with combinations of powers that do not appear in $f(\nu)$ and vice versa.
We reluctantly conclude that the quantum imprint rule is an exact result strictly in the limit $\nu \to \infty$ and $\hbar \to 0$.
Nevertheless, as an exact result, the quantum imprint rule can be interpreted as a quantum anomaly, similar to the Adler-Bell-Jackiw anomaly~\cite{Bell:1969ts, Adler:1969gk} for axial $U(1)$ symmetries.
Specifically, the classical $\mathbb{Z}_2$ symmetry of the theory is broken by $f(\nu)$ upon quantization, and the breaking enters the matrix elements of the theory by means of the quantum imprint.
In other words, the quantum imprint violates the semiclassical limit in a qualitatively similar way that instantons explicitly violate amplitudes that measure classical conservation laws~\cite{tHooft:1976snw}.

We remark that the need of a double scaling limit to uncover the quantum imprint rule coincides with general arguments against strict reductionism~\cite{Berry1995597,Batterman2002}.
Indeed, the limit $\hbar \to 0$ is singular, and thus the naive correspondence principle fails, allowing the possibility of emergent phenomena.\footnote{For another striking example where the correspondence principle fails, consider two on-phase particles of de~Broglie wavelength $\lambda_{\mathrm{DB}}$ colliding along the $x$-axis. By linearity, the total intensity is proportional to $\cos^2 \left( x / \lambda_{\mathrm{DB}} \right)$. In this case, the correspondence principle limit $\lambda_{\mathrm{DB}} \to 0$ is singular and does not approach the classical result, which is simply twice the intensity of either particle.}

\subsection{Other quantum imprint examples}
\label{subsec:universality}

Thus far, we have derived the quantum imprint rule for the anharmonic oscillator featuring a symmetric double well potential.
We note that our arguments can be repeated almost verbatim for the anharmonic single well potential,
\begin{equation}
	V_{\mathrm{SW}}(x) = \frac{1}{32g^2} \left( 1 + 4g^2x^2\right)^2 \, ,
\end{equation}
albeit with a different explicit form of $f(\nu)$ and $\braket{\nu|x|0}$.
In this case, the classically forbidden region characterized by the complex contour $\mathcal{B}$ is found on the imaginary axis for any positive value of the energy. 
Here, in contrast to the double well, the perturbative series expansion of the energy levels is tractable and Borel-Laplace resummable to the exact result~\cite{Simon:1970mc, Graffi:1970erh}.\footnote{In some cases, summability to an exact result also applies to higher-dimensional QFTs~\cite{10.1007/BF01705374, Magnen:1977ha}. A more general, systematic assessment of Borel-Laplace summability can be found in~\cite{Serone:2017nmd}.}
Nonetheless, an exact WKB treatment of the theory is also interesting to consider, and must include both real and imaginary instantons, since the corresponding perturbative quantum period is unfortunately not Borel-Laplace summable~\cite{10.1103/PhysRevLett.41.1141, Grassi:2014cla}.
We have performed this exact WKB calculation explicitly and our results indeed show that \cref{eq:NPHolyGrail} also holds for the anharmonic oscillator featuring a single well potential and hence reinforce the quantum imprint rule.

Beyond this example, we can outline possible avenues to explore how to generalize the quantum imprint rule~\eqref{eq:NPHolyGrail}.
One crucial aspect to analyze is the classical vacuum structure of the theory, which is $\mathbb{Z}_2$ symmetric in the anharmonic double well potential and hence possesses four turning points, \textit{i.e.}~two independent cycles.
The underlying geometry is thus that of a genus-one surface, which allowed us in~\cref{ssec:PertQuant} to use \textit{e.g.} powerful cohomology results~\cite{Fischbach:2018yiu, Kreshchuk:2018qpf}.
As an aside, it is worth noting that the geometry of higher genera surfaces is still poorly understood, and thus theories with more complicated vacuum structures are likely intractable to calculate at present.
Nevertheless, we conjecture that a modified version of~\cref{eq:NPHolyGrail} may still hold in theories with more complicated vacuum degeneracies, where the modification tallies the leading instanton amplitudes.

We end this discussion by noting the quantum imprint rule is a concrete realization of a perturbative-nonperturbative connection as predicted by resurgence theory (see, \textit{e.g.}, Refs.~\cite{Zinn-Justin:2004vcw, Zinn-Justin:2004qzw, Jentschura:2004jg, Dunne:2014bca}).
Indeed, the amplitude $\braket{\nu | x | 0}$ was calculated using perturbation theory alone, while $f(\nu)$ is a purely nonperturbative quantity.
In this sense, \cref{eq:NPHolyGrail} furnishes an explicit breaking of the factorization between the perturbative and nonperturbative sectors.

\section{Conclusions and Outlook}
\label{sec:conclusions}

In this work, we have employed an exact WKB analysis to identify purely perturbative and nonperturbative sectors of the anharmonic oscillator featuring a symmetric double well potential.
The nonperturbative contribution to the quantization condition can be entirely characterized by the function $f(\nu)$, which measures the tunneling through the classically forbidden potential barrier that separates the minima and also generates energy splitting between classically degenerate levels.
Remarkably, at large quantum numbers, or more precisely in the classical double scaling limit $\nu \to \infty$ and $\hbar \to 0$ with $\hbar \nu$ fixed, we have found that $f(\nu)$ is precisely given by transition amplitudes to highly excited states, $\braket{\nu | x | 0}$, as shown in \cref{eq:NPHolyGrail}.
The transition amplitude is the quantum mechanical analogue of multiparticle production in scalar QFTs at high energies, which becomes nonperturbative in the limit of a large number of final state particles.

Na\"{i}vely, the quantum states at large $\nu$ are so energetic that they should not feel the potential barrier of the double well.
Nevertheless, the right hand side of \cref{eq:NPHolyGrail} is given by the instanton action.
We interpret this as the irreducibility of the quantum nature of the theory, even in the correspondence principle limit.
In other words, the instanton characterizes the leading deviation of the theory from classical behavior.
Being an all-orders result, it gaps the quantum theory, as there is no value of the coupling where this contributions vanishes entirely.
Therefore, strictly speaking, the theory does not exhibit a classical regime and is instead intrinsically quantum, separated from its classical counterpart by the instanton action, and we hence dub \cref{eq:NPHolyGrail} the quantum imprint rule.

Furthermore, we have found that the same quantum imprint rule also holds in the case of an anharmonic oscillator with a single well potential.
This leads us to speculate that a general version of the quantum imprint rule can be derived in any quantum mechanical theory, provided the right hand side is modified to account for different vacuum structures, which we plan to address in future work.
We also conjecture that a more general form of~\cref{eq:NPHolyGrail} can be used to define universality classes of different theories via the quantitative measure that gaps the theory from its classical behavior.
In this sense, we can highlight that the double well and the single well anharmonic oscillators, while maintaing superficial differences in the explicit form of $f(\nu)$ and $\braket{\nu|x|0}$, are similar to each other in the semiclassical limit because they belong to the same universality class governed by~\cref{eq:NPHolyGrail}.

Lastly, it would be very interesting to derive an analogous identity to \cref{eq:NPHolyGrail} in QFT.
Namely, the analogy between quantum anomalies and \cref{eq:NPHolyGrail} is striking since the latter violates the classical limit in a qualitatively similar way that amplitudes measuring anomalous classical conservation laws are violated explicitly by instantons.
We can also offer \cref{eq:NPHolyGrail} as a measure to distinguish universality classes of QFTs apart from their respective classical limits.
In the event that the quantum imprint rule is realized as an exact relation in QFT, its application as an operative equation to derive perturbative or nonperturbative physics amplitudes would be highly effective.

\section*{Acknowledgments}

The authors would like to thank the Fermilab theory group for their kind hospitality while this work was in progress. 
We thank Marcos Mari\~no for very insightful discussions.
FY would like to acknowledge insightful conversations with Daniel Chung and Patrick Draper.
PLC would like to acknowledge insightful conversations with Roni Harnik.
This work is supported by the Cluster of Excellence {\em Precision Physics, Fundamental Interactions and Structure of Matter} (PRISMA${}^+$ -- EXC~2118/1) within the German Excellence Strategy (project ID 39083149).

\appendix

\section{Asymptotic Series and Borel-Laplace Summation}
\label{app:BorelLaplace}

In this appendix, we review Borel-Laplace summation with an explicit example, closely following the presentation in Ref.~\cite{SauzinSummability}.

A partial series $\phi = \sum_{i=0}^{N} a_i x^i$ is said to be asymptotic to a function $f(x)$ if, for fixed $N$ and $x \ll 1$, its partial sum approximates $f$ to polynomial accuracy,
\begin{equation}
    \abs{f(x) - \phi(x;N)} \sim \mathcal{O} \left(x^{N+1} \right) \, .
\end{equation}
The first thing to note is that ``being asymptotic to" is not a one-to-one relation, but one-to-many.
Indeed, consider the function $\exp (-1/x)$, whose Taylor expansion at zero is trivial, given that for all terms, at each order $n$,
\begin{equation}
    \left.\frac{\md^n}{\md x^n} \me^{-\frac{1}{x}} \right\rvert_{x=0} = 0 \, .
\end{equation}
Hence, if $\phi$ is asymptotic to $f(x)$ around $x =0$, then it is also asymptotic to $f(x) + \exp(-1/x)$.
The second thing to note is that the definition of an asymptotic series tells us nothing about the limit for fixed $x$, $N \to \infty$ which is used to assess convergence of $\phi$ to $f(x)$ along the whole range of $x$.
Asymptotic series indeed can be divergent, and even display zero radius of convergence.
Nonetheless, an asymptotic divergent series is not devoid of information and together with a resummation procedure it can be used -- in favourable cases -- to reproduce the exact answer to the problem at hand.

Borel-Laplace summation is perhaps the simplest and best known resummation method.
Given a formal series $\phi$, we define its Borel-Laplace summation as the function $\hat\phi = \mathcal{L}\mathcal{B}\phi$.
Here, we have defined the Borel transform $\mathcal{B}$,
\begin{equation}
    \phi = \sum_{n=0}^{\infty} a_n z^{n} \to \mathcal{B}\left( \phi \right)= \sum_{n=0}^{\infty} \frac{a_n}{t^{n+1}n!}\, ,
\end{equation}
defining a map between series defined around zero ($z\ll 1$) and series defined around infinity ($t \gg 1$), or vice versa, and the Laplace transform $\mathcal{L}$,
\begin{equation}
    \mathcal{L}\phi(z) = \int_0^\infty \md t \, \me^{-zt} \phi(t) \, .
\end{equation}

As an example of this resummation procedure, consider the equation
\begin{equation}
    \frac{\md \Phi}{\md z}=\Phi - \frac{1}{z} \, ,
\label{eq:Euler}
\end{equation}
that admits the Euler series $\Phi = \sum_n (-1)^n \, n! \, z^{-(n+1)}$ with $z \gg 1$ as a formal solution.
The series $\Phi$ clearly exhibits factorial growth of its coefficients and hence has zero radius of convergence.
The Borel transform of $\Phi$ is the geometric series
\begin{equation}
    \mathcal{B}(\Phi) =\sum_{n=0}^{\infty}(-1)^nt^n = \frac{1}{1+t}\, ,
\end{equation}
and the pole at $t = -1$ is a consequence of the divergence of the original Euler series.
The Laplace transform of the geometric series gives a particular solution of \cref{eq:Euler} with finite behavior at infinity,
\begin{equation}
    \mathcal{L}(1+t)^{-1} = \me^z \, \Gamma(0;z)\,,
\end{equation}
where $\Gamma(a;z) = \int_z^\infty \md t \, \me^{-t}\, t^{a-1}$ is the upper incomplete gamma function.

In this simple example, the basic motivation behind why Borel-Laplace resummation works is evident: namely, it relies on the interchange between summation and integration using the integral representation of the Gamma function
\begin{equation}
    \sum_{n=0}^{\infty} (-1)^{n} \, n! \, x^{n} = \sum_{n}^{\infty} \left( -1 \right)^n
    \left( \int_0^\infty \md t \, \me^{-t} \, t^{n} \right) \, x^{n}
    = \int_{0}^{\infty} \md t \, \me^{-t} \frac{1}{1+xt} \, .
\end{equation}
It is also easy to see when this procedure fails.
If $\mathcal{B}(\phi)$ has poles along the real axis, the Laplace transform cannot be performed.
For instance, consider the absolute value of the Euler series, $\tilde{\Phi}=\sum_{n} n! \, z^{-(n+1)}$.
Its Borel-Laplace transformation is
\begin{equation}
    \mathcal{L}\mathcal{B} \, \tilde{\Phi} = \int_0^\infty \md t \, \frac{ \me^{-zt}}{1-t} \, ,
\end{equation}
which is not well-defined because the integrand has a pole at $t=1$.
However, we can consider slight deformations by $\epsilon$ on the positive or negative imaginary axis, such that
\begin{equation}
    \mathcal{L}^{\pm \epsilon}\mathcal{B}\tilde{\Phi} \approx \int_0^\infty \md t \, \frac{\me^{-zt}}{1-t\pm i\epsilon} \, .
\end{equation}
Both of these integrals converge but not to the same value,
\begin{equation}
   \mathcal{L}^{+ \epsilon}\mathcal{B}\tilde{\Phi}-\mathcal{L}^{-\epsilon}\mathcal{B}\tilde{\Phi}= 2\pi i \me^{-z} \, .
\label{eq:Ambiguity}
\end{equation}
Series that display such behavior are said to be non Borel-Laplace summable.
The theory of resurgence provides a prescription for working with such series and resolves the ambiguity in \cref{eq:Ambiguity}.

\bibliographystyle{inspire}
\bibliography{refs, refs_noninspire}

\end{document}